\documentclass[a4paper,10pt]{aa}
\usepackage{graphicx}
\newcommand{\halfskip}{\vskip 0.5\baselineskip \noindent}
\newcommand{\be}{\halfskip \begin{equation}}
\newcommand{\ee}{\end{equation} \halfskip}

\begin{document}
\authorrunning{E. van der Swaluw et al.}
\title{An evolutionary model for pulsar-driven supernova remnants}
\subtitle{a hydrodynamical model}
\author{E. van der Swaluw
\inst{1,2},
T.P. Downes
\inst{3},
\and
R. Keegan
\inst{3}}
\institute{Dublin Institute for Advanced Studies, 5 Merrion Square, Dublin 2, 
Ireland
\and
FOM-Institute for Plasma Physics Rijnhuizen, P.O. Box 1207, 3430 BE Nieuwegein, The Netherlands 
\and
School of Mathematical Sciences, Dublin City University, Glasnevin, Dublin 9,
Ireland
}
\offprints{E. van der Swaluw,
\email{swaluw@rijnh.nl}}
\date{}
\abstract{We present a model of a pulsar wind nebula evolving inside its 
associated supernova remnant. The model uses a hydrodynamics code to simulate
the evolution of this system when the pulsar has a high velocity.
The simulation distinguishes four different stages of pulsar wind nebula 
evolution: the supersonic expansion stage, the reverse shock interaction
stage, 
the subsonic expansion stage and ultimately the bow shock stage.
The simulation bears out that, due to the high velocity of the pulsar, the 
position of the pulsar is off-centered with respect to its 
pulsar wind nebula, after the passage of the reverse shock. Subsequently the pulsar 
wind nebula expands subsonically untill the event of the bow shock formation, when 
the motion of the pulsar becomes supersonic. The bow shock formation event occurs 
at roughly half the crossing time, when the pulsar is positioned at 0.677 times 
the radius of the supernova remnant blastwave, in complete agreement with analytical
predictions. The crossing time is defined by the age of the supernova remnant, when 
the pulsar overtakes the blastwave bounding the supernova remnant.  

The results of the model are applied to three supernova remnants: N157B,
G327.1-1.1 and W44. We argue that the head of the pulsar wind nebula,
containing the active pulsar, inside the first two systems are not bounded 
by a bow shock. However, in the case of W44 we argue for a scenario in which 
the pulsar wind nebula is bounded by a  bow shock, due to the supersonic motion 
of the pulsar.
\keywords{Pulsars -- Supernova Remnants -- Shocks -- Hydrodynamics}}
\maketitle

\section{Introduction}
A supernova remnant (SNR) is the relic of a supernova explosion, which 
injects an energy of $\sim 10^{51}$ erg into the surrounding medium. 
The dynamics of a young SNR expanding into the interstellar medium (ISM) 
is determined by two shocks: a forward shock which propagates into the ISM 
and is being decelerated by sweeping up material from the ISM, and a reverse 
shock which results from the high pressure behind the forward shock and 
propagates back into the freely expanding ejecta of the SNR (McKee 1974; 
McKee \& Truelove 1995).

A young SNR becomes dynamically more interesting in those cases where the 
collapse of the progenitor star, preceding the supernova explosion, yields
a pulsar: a rapidly rotating neutron star. In those cases the dynamics of the
central region of the SNR is dominated by the continuous injection of 
energetic particles by a relativistic pulsar wind, driven by the spin-down 
energy of the pulsar. The pulsar wind is terminated by a strong MHD shock 
(Rees \& Gunn 1974), and drives a pulsar wind nebula (PWN) in the interior 
of the young SNR. The dynamics of the PWN is coupled to the evolution of 
the SNR, because the total energy release over the pulsar's lifetime is small 
($\sim 10^{49}-10^{50}$ erg) compared with the total mechanical energy of the 
SNR ($\sim 10^{51}$ erg). 

Several authors (Reynolds \& Chevalier 1984; van der Swaluw et al. 2001;
Blondin et al. 2001; Bucciantini et al. 2003) have considered the 
evolution of a centered PWN inside an evolving SNR. 
In these systems, the initial stage of the PWN is the supersonic expansion
stage: the pulsar wind bubble is bounded by a strong PWN shock propagating 
through the freely expanding ejecta of the SNR. A transition to the reverse
shock interaction stage takes place when the reverse shock collides with the 
PWN shock, and subsequently crushes the pulsar wind bubble 
(van der Swaluw et al. 2001, Blondin et al. 2001). 
This reverse shock interaction stage is characterised by an unsteady expansion 
of the pulsar wind bubble (van der Swaluw et al. 2001), due to the reverberations 
from the violent collision between the reverse shock and the PWN shock.
The expansion of the PWN
proceeds subsonically when these reverberations have vanished. The expansion is
subsonic because the surroundings of the PWN have been reheated by the passage of
the reverse shock: the PWN shock, bounding the swept-up shocked ejecta around the 
hot pulsar wind bubble, has disappeared.

In this paper we discuss the evolution of PWNe inside SNRs, for pulsars
with a constant kick velocity: initially the PWN starts its expansion at
the center of the SNR, however the pulsar motion will move the pulsar wind 
cavity along 
as it moves through the SNR interior. We present results from hydrodynamical 
simulations for such a system, which distinguishes all three evolutionary stages
mentioned above, i.e. the supersonic expansion stage, the reverse shock 
interaction stage and the subsonic expansion stage. However at the end of the simulation 
an additional stage can be distinguished, when the head of the PWN, containing
the active pulsar, deforms into a bow shock, due to the supersonic motion of
the pulsar.

The supersonic expansion stage in the simulation shows a PWN which is
off-centered with respect to the twofold shock structure of the SNR,
due to the kick velocity of the pulsar. Therefore the timescale on which
the reverse shock collides with the complete shock surface, bounding the 
PWN, can be a significant fraction of the total lifetime of the PWN when
the reverse shock interaction stage starts. We use a semi-analytical approach 
to estimate the timescale of the collision process, which is shown to scale 
roughy with the pulsar velocity. The results from the simulation are in almost
complete agreement with these semi-analytical calculations.
Due to the high velocity of the pulsar, its position inside the PWN is strongly 
off-centered after the passage of the reverse shock. Ultimately, at the end of 
the simulation, when the pulsar is approaching the shell of its SNR, the head of 
the PWN, containing the active pulsar, is deformed into a bow shock at a time and 
position, in complete agreement with analytical predictions made by van der Swaluw 
et al. (1998).

We determine the evolutionary stage of the PWNe inside the SNRs N157B, G327.1-1.1 
and W44, using our model. For the first two systems, we argue that the position of
the pulsar at the head of its PWN, 
is a result of the passage of the reverse shock and the high velocity of the pulsar:
the head of these PWNe are not bounded by a bow shock.
Therefore these PWNe are either in the 
{\it reverse shock interaction stage} or the {\it subsonic expansion stage}. The PWN 
inside the SNR W44 however, is shown to be a good candidate for having a bow shock nebula 
around its pulsar.

\section{The evolution of a PWN inside a SNR}

\subsection{The interaction between the reverse shock and the PWN shock}

The initial stage of PWN evolution is characterised by a hot pulsar wind bubble,
bounded by a strong PWN shock, propagating through the freely expanding ejecta
of the SNR. A transition to the subsonic expansion stage occurs via the 
reverse shock interaction stage. This interaction stage starts when the reverse shock collides with
the PWN shock. In the case of a centered pulsar ($V_{\rm psr}=0$), the reverse
shock collides with the PWN shock surface instantaneously, due to the
spherical symmetry of the SNR and the PWN. However, when the pulsar has a kick 
velocity, there will be an associated timescale on which the reverse shock collides 
with the complete surface of the PWN shock. This is the first stage of the 
reverse shock interaction stage. Next, the pulsar wind bubble oscillates back and forward 
due to the presence of reverberations from the passage of the reverse shock (van der Swaluw
et al. 2001). The reverse
shock interaction stage ends when these reverberations have vanished and the pulsar
wind bubble proceeds its expansion subsonically. In this section we use a 
semi-analytical approach to derive a timescale on which the reverse shock 
collides with the complete surface of the PWN shock.

McKee \& Truelove (1995) give analytical approximations for the trajectories of the
forward shock and the reverse shock of a SNR in the case of a uniform ambient medium. Their
equations for the trajectories of both shocks are normalised to a timescale $t_{\rm ST}$,
which marks the age of the remnant when it has swept up roughly 1.61 the ejected mass 
$M_{\rm ej}$. Their equations descibe the expansion of an isolated SNR in the free
expansion stage and the Sedov-Taylor stage. The trajectory of the forward shock
converges to the Sedov-Taylor solution when the SNR age $t >> t_{\rm ST}$. Internal
(radiative) cooling is not inlcuded in their model, therefore the pressure-driven
snowplow stage is not described (see however Blondin et al. 1998).

\begin{figure}
\resizebox{\hsize}{!}
{\includegraphics{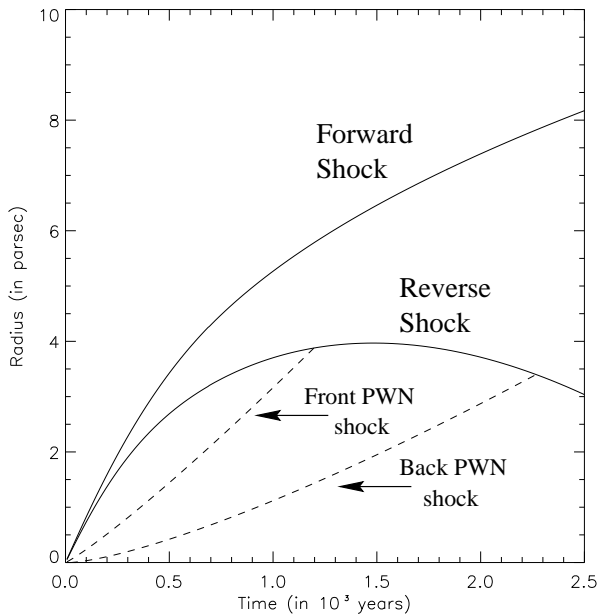}}
\caption{Radii of the forward and the reverse shocks of a SNR with an
explosion energy of $E_0=10^{51}$ erg and a total ejected mass of 
$M_{\rm ej}=3\; M_\odot$.
The dashed lines indicate the position of the front ($R^{\rm f}_{\rm pwn}$) 
and the back ($R^{\rm b}_{\rm pwn}$) of the PWN shock. The luminosity of the 
pulsar wind is constant, $L_0=10^{38}$ ergs/sec and the velocity of the 
pulsar equals $V_{\rm psr} = 1\;000$ km/sec. The timescale on
which the reverse shock collides with the complete shock surface of the 
PWN equals $\Delta t_{\rm col}\simeq 1\;050$ years. This timescale is of the same
order as the age of the PWN, $t\simeq 1\;200$ years, on which the reverse shock 
interaction stage starts.}
\end{figure}

The reverse shock hits the center of the SNR at approximately 5 times the 
transition time $t_{\rm ST}$, putting an upper limit on the age $t_{\rm col}$, when 
the reverse shock has collided with the complete shock surface bounding the PWN:
\be
t_{\rm col}\;\simeq\; 5t_{\rm ST}\; =\;  1\; 045 E^{-1/2}_{51}
\left({M_{\rm ej}\over M_\odot}\right)^{5/6}n_0^{-1/3}\;\;{\rm years}\; ,
\ee
here $E_{51}$ is the total mechanical energy of the SNR in units of $10^{51}$ erg 
and $n_0$ is the ambient hydrogen number density assuming an interstellar composition
of 10 H : 1 He. The above timescale is very close to the one given by Reynolds\& Chevalier 
(1984).

An equation for the radius of the PWN shock, when it is
interacting with the freely expanding ejecta of its SNR was given by van der 
Swaluw et al. (2001), where a constant pulsar wind luminosity $L_{\rm pw}=L_0$ was taken:
\be
\label{Anpwn}
        R_{\rm pwn}(t)\;\simeq\; 0.889\left({L_0t\over E_0}\right)^{1/5}V_0t\propto t^{6/5},
\ee
here $E_0$ is the total mechanical energy of the SNR and $V_0$ is defined as:
\be
V_0\; =\; \sqrt{{10\over 3}{E_0\over M_{\rm ej}}} \; .
\ee
We have taken the case for which the adiabatic heat ratio of the pulsar wind 
material equals $\gamma_{\rm pwn}=5/3$. 

We will use the equations for the trajectories of the forward and the reverse shock 
of the SNR from Truelove 
\& McKee (1995) and the above equation for the radius of the PWN shock to 
calculate the collision time $t_{\rm col}$, i.e. when the reverse shock 
collides with the PWN shock. We make a distinction between 
the front of the PWN shock $R^{\rm f}_{\rm pwn}=R_{\rm pwn}+V_{\rm psr}t$, 
and the back of the PWN shock $R^{\rm b}_{\rm pwn}=R_{\rm pwn}-V_{\rm psr}t$,
which enables us to calculate $\Delta t_{\rm col}$, the timescale on which
the reverse shock collides with the complete shock surface bounding
the PWN. 
This approach neglects the change in the PWN expansion due to the displacement
of the pulsar inside the freely expanding ejecta.
Figure 1 shows an example of this approach, where the position of the four different
shocks are plotted as a function of time. The parameters have been taken 
similarly to the parameters which we will use in the simulation. The figure
shows that the timescale on which the reverse shock collides with the complete
PWN shock surface equals  $\Delta t_{\rm col}\simeq$ 1050 years.
Figure 2 shows $\Delta t_{\rm col}$ as a function of the pulsar velocity,
where one observes that this timescale correlates almost linearly with the pulsar 
velocity. We conclude that the collision timescale $\Delta t_{\rm col}$ of PWNe 
containing a high velocity pulsar, can be a significant fraction of the timescale 
associated with the supersonic expansion stage.   

\begin{figure}
\resizebox{\hsize}{!}
{\includegraphics{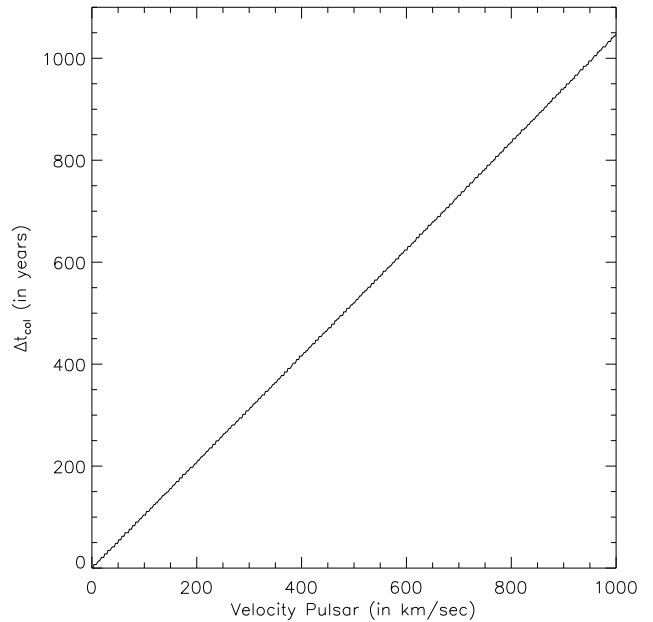}}
\caption{The timescale $\Delta t_{\rm col}$ on which the reverse shock collides
with the complete shock surface bounding the PWN, as a function of the pulsar
velocity. Apart from the velocity, the same parameters have been taken as in 
figure 1.}
\end{figure}

\subsection{The formation of the PWN bow shock}
After the passage of the reverse shock 
the PWN consists of two distinct parts:
\begin{itemize}
\item
The roughly spherically symmetric PWN relic, from the initial energetic stage of 
the pulsar wind, which will ultimately expand subsonically, after the reverberations
of the reverse shock have vanished.
\item
The relic PWN connects with the head of the PWN, directed towards the shell of the 
remnant. The head of the PWN contains the active pulsar, which propagates towards 
the SNR shell.
\end{itemize}
The age of the remnant when the pulsar overtakes the shell of its remnant was
given by van der Swaluw et al. (2003):
\be
\label{crosstime}
       t_{\rm cr}\;\simeq\; 1.4 \times 10^4 \; E_{51}^{1/3} 
       V_{\rm 1000}^{-5/3}n_0^{-1/3}\; {\rm years} .
\ee       
$V_{\rm 1000}$ denotes the velocity of the pulsar in units of 1000 km/sec.
This timescale was calculated in the limit of a Sedov-Taylor SNR, i.e. it is 
assumed that the pulsar will break through the shell of the remnant, when 
radiative losses of the SNR are neglible.  

We follow van der Swaluw et al. (1998) by calculating the Mach number 
${\cal M}_{\rm psr}$ of the pulsar as it
propagates through the SNR interior, using the Sedov-Taylor solution 
(Sedov 1959). Figure 3 shows the Mach number ${\cal M}_{\rm psr}$ as a function 
of the age of the remnant $t$, normalised to the crossing time $t_{\rm cr}$. 
Alternatively Figure 4 shows the  Mach number ${\cal M}_{\rm psr}$ as a function 
of the position of the pulsar $R_{\rm psr}$, normalised to the position of the SNR 
blastwave $R_{\rm snr}$. 

\begin{figure}
\resizebox{\hsize}{!}
{\includegraphics{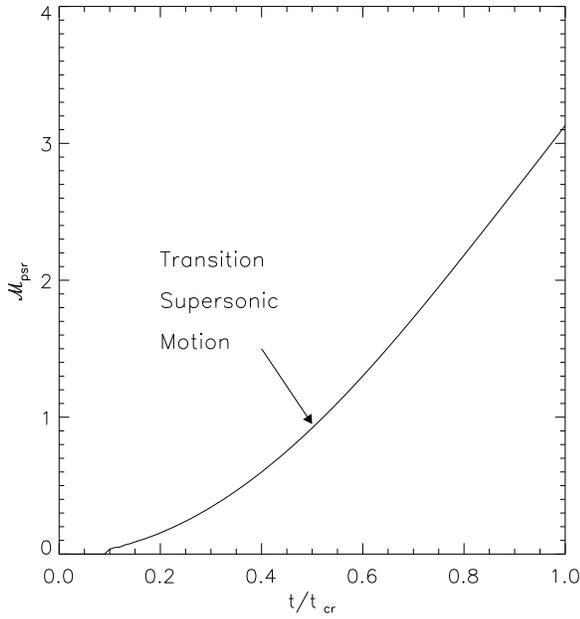}}
\caption{The Mach number of the pulsar ${\cal M}_{\rm psr}$, as a function of 
the SNR age $t$ scaled to the crossing time $t_{\rm cr}$ of the pulsar. The
event of bow shock formation (${\cal M}_{\rm psr}=1$) occurs at $t=0.5t_{\rm cr}$.}
\end{figure}

The figures show that the PWN head, containing the active pulsar, is deformed into a 
bow shock (i.e. when ${\cal M}_{\rm psr}=1$), at half the crossing time when the pulsar 
is positioned at a radius of $0.677$ times the radius of the SNR blastwave.  
The Mach number will slowly increase after the formation of the bow shock, due to the 
decrease of the sound speed, as the PWN is approaching the shell of the remnant. When 
the pulsar breaks through the shell, the Mach number equals 
${\cal M}_{\rm psr}={7\over\sqrt{5}}\simeq 3.13$ (van der Swaluw et al. 2003). 

A lower limit for the pulsar velocity can be derived, such that the bow shock
formation occurs while the SNR is in the Sedov-Taylor stage.
We follow van der Swaluw et al. (2003), who 
already give a lower limit for the pulsar velocity in order for the pulsar to 
cross the SNR shell in its Sedov-Taylor stage. They use the transition time 
calculated by Blondin et al. (1998), as the age of the SNR when the transition 
from the Sedov-Taylor stage to the pressure-driven snowplow stage occurs:
\be
t_{\rm PDS} = 2.9 \times 10^4 E_{51}^{4/17}n_0^{-9/17} \; {\rm yr}.
\ee
Our results show that the bow shock creation occurs at half the crossing time,
therefore the requirement $0.5t_{\rm cr}\le t_{\rm PDS}$ yields a lower limit
for the pulsar velocity to let the bow shock formation event occur when the SNR 
is in its Sedov-Taylor stage:
\be       
       V_{\rm psr} \ge 325\; n_0^{2/17} E_{51}^{1/17}
       \; {\rm km/s} \; ,
\ee
which is a reasonable fraction of the observed pulsar velocities by Hansen and
Phinney (1997). 

\begin{figure}
\resizebox{\hsize}{!}
{\includegraphics{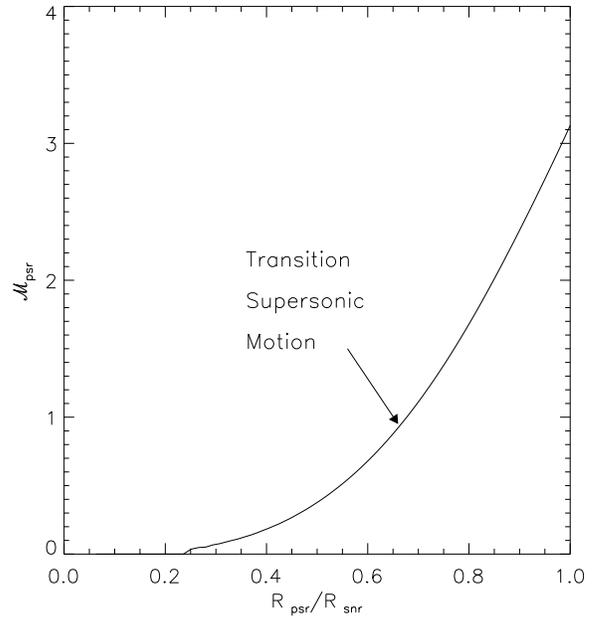}}
\caption{The Mach number of the pulsar ${\cal M}_{\rm psr}$ as a function of the 
ratio of the pulsar position $R_{\rm psr}$ and the SNR blastwave $R_{\rm snr}$.
The bow shock formation  of the SNR. The event of the bow shock formation 
(${\cal M}_{\rm psr}=1$) occurs at $R_{\rm psr}/R_{\rm snr}\simeq 0.677$.}
\end{figure}

\section{Numerical Simulations}

\subsection{Simulation method}

We use a second order, properly upwinded hydrodynamics code (described
in Downes \& Ray,\cite{d&r99}) to simulate the dynamics of the
interaction between a pulsar wind and a supernova remnant when the pulsar
has a high velocity.  
The hydrodynamics equations are intergrated in cylindrical symmetry, and the
boundary conditions are taken as gradient zero everywhere except on the
$r=0$ boundary, where they are set to reflecting.


\begin{figure*}
\centering
\includegraphics[width=12cm,angle=0]{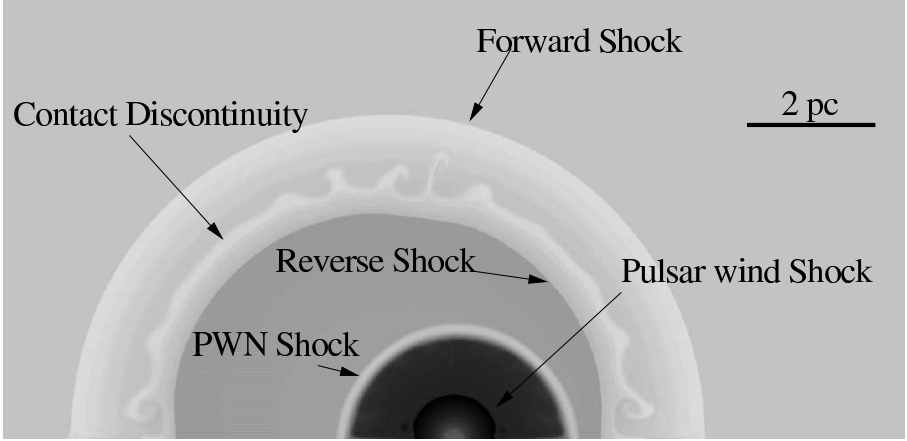}
\caption{Logarithmic gray-scale plot of the density distribution at an age
$t_{\rm snr}$ = 1 000 years.}
\end{figure*}


\subsection{Initialising the system}

The simulations are performed in the rest frame of the pulsar. The pulsar
velocity equals $V_{\rm psr}=1\;000$ km/sec.  
Initially, the ambient density and pressure are uniform with a uniform
velocity of $1\;000$ km s$^{-1}$ in the negative $z$ direction.  The ambient
density is $\rho_0=10^{-24}$ g/cm$^{-3}$ (corresponding with $n_0\simeq 0.427$), 
and the pressure is chosen so that the ambient temperature is 
$T = 1\times 10^5$ K.  The resolution is taken as 
$\Delta r =\Delta z = 3.6\times10^{16}$ cm.

An exanding SNR is created by initialising a sphere of radius 0.25 pc, or 
21 grid cells, with a high thermal energy and density such that the total 
energy contained in the sphere is $E_0=10^{51}$ ergs, while the ejecta mass is 
$M_{\rm ej}=3 M_{\odot}$.  Note that this material also has a velocity of 
$1\;000$ km s$^{-1}$ in the negative $z$ direction.

The source of the pulsar wind is modeled as a small
sphere of radius 0.175 pc (15 grid cells) into which thermal energy
is deposited at a rate of $L_{\rm pw}=10^{38}$ erg/s. Mass is also deposited into 
this region at a rate chosen such that the pulsar wind terminal velocity is
$v_{\infty}=30\;000$ km s$^{-1}$. The pulsar wind luminosity has been taken constant, in order
to resolve the pulsar wind termination shock throughout the simulation. The results
of the simulation will not change qualitatively, because the total injected energy 
by the pulsar during its stay in the SNR interior 
$E_{\rm pw}={L_{\rm pw} t_{\rm cr}}\ll E_0$.

\begin{figure*}
\centering
\includegraphics[width=12cm,angle=0]{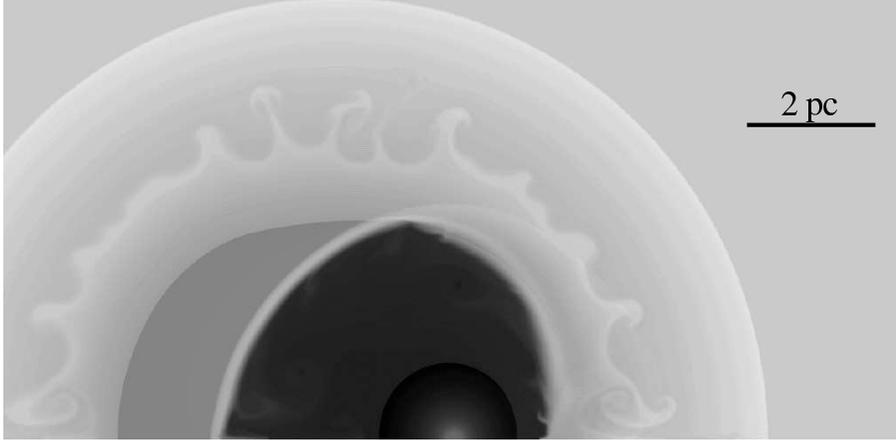}
\caption{Logarithmic gray-scale plot of the density distribution at an age
$t_{\rm snr}$ = 1 800 years.}
\end{figure*}

%

\subsection{The reverse shock interaction stage}

The initial stage of the PWN evolution is the supersonic expansion stage. Figure 5 
shows the density profile of the PWN/SNR system during this 
stage. One can clearly distinguish the four different shocks of the system: the 
pulsar wind termination shock, the PWN shock, the reverse shock and the forward shock.   
One clearly observes the off-centered position of the PWN with respect to the 
twofold shock structure of the SNR, caused by the motion of the pulsar. Furthermore
the pulsar wind cavity is roughly spherically symmetric. This results from the high
sound speed inside the pulsar wind bubble, which smooths out pressure perturbations 
rapidly, yielding an almost uniform pressure distribution in the PWN.

Figure 6 shows the density profile of the PWN/SNR system, 
when the front of the PWN shock has already collided with the reverse shock. At this
stage the pulsar position starts to get off-centered with respect to the PWN structure.
This is caused by the reverse shock interaction, which raises the presssure inside the
disturbed parts of the bubble, which results in pulsar wind material flowing towards 
the back of the PWN. Furthermore the increase in the pressure inside the front part of 
the pulsar wind bubble, pushes the forward termination shock backwards (towards the pulsar)
as well. On the other hand, the downstream pressure at the backward termination shock 
has not been influenced by the reverse shock interaction yet. Therefore the pulsar wind 
cavity structure is no longer spherically symmetric as can be seen in Figure 6.
The total time on which the reverse shock collides with the {\it complete} shock structure bounding
the pulsar wind bubble equals $\Delta t_{\rm col}\simeq 1\;200$ years in the simulation, after which 
the age of the SNR equals $t \simeq 2\;400$ years. These timescales are in almost complete agreement 
with the analysis performed in section 2.1 of this paper, which validates the approach made
in that section, to derive the collision timescale $\Delta t_{\rm col}$. 

%

Figure 7 shows the PWN/SNR system shortly after the passage of the reverse shock.
The pulsar is positioned at the head of the PWN, due to the passage of the reverse shock and
the high velocity of the pulsar. The PWN structure consists of roughly two parts, a relic
PWN and the head of the PWN, as was described in section 2.2 of this paper.
After the passage of the reverse shock, the reverse shock interaction stage is continued by the 
reflected and transmitted shock waves, which propagate through both the pulsar wind bubble 
and the surrounding ejecta of the SNR. This stage was described in detail by van der Swaluw 
et al. (2001) for a spherically symmetric PWN/SNR system. We observe a qualitatively similar 
process, in which the pulsar wind bubble is oscillating. The dynamics are more complicated
compared with the simulations of van der Swaluw et al. (2001), due to the asymmetry of the 
PWN/SNR system caused by the motion of the pulsar. During this stage of the reverse shock 
interaction, the PWN is oscillating between expansion and compression and the simulations 
reveal Rayleigh-Taylor and Kelvin-Helmholtz instabilities. These instabilities lead to the 
mixing of ejecta and pulsar wind material, as was similarly observed in the simulations 
performed by Blondin et al. (2001). Therefore at the end of the reverse shock interaction 
stage, the relic PWN consists of a mixture of pulsar wind material and ejecta. In our 
simulation the relic PWN will move off the grid, before the bow shock formation occurs around 
the head of the PWN.

\subsection{The formation of the bow shock}

At the end of the simulation, as the pulsar approaches the shell of the SNR, the
head of the PWN, containing the active pulsar, deforms into a bow shock, due to the
supersonic motion of the pulsar. The simulation shows that the bow shock formation
event occurs at roughly $t\simeq 0.5 t_{\rm cr}$, when the position 
of the pulsar $R_{\rm psr}$ with respect to the radius of the blastwave $R_{\rm snr}$
equals $R_{\rm psr}/R_{\rm snr}\simeq 0.677$. This clearly demonstrates the
validity of the analytical approximation made in section 2.2 of this paper.
Figure 8 shows the density profile of the PWN/SNR system after the bow shock formation. The 
curvature of the SNR shock is small compared with the bow shock structure, which validates 
the assumptions made by van der Swaluw et al. (2003) to model the break-through event.

\begin{figure*}
\centering
\includegraphics[width=12cm,angle=0]{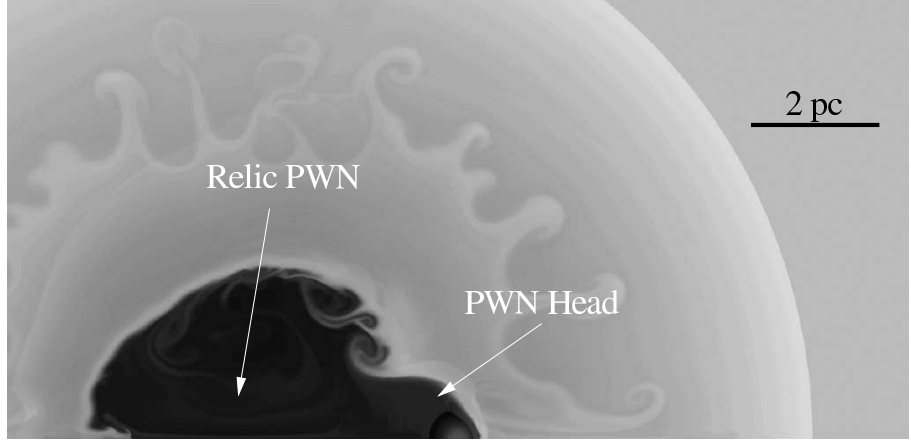}
\caption{Logarithmic gray-scale plot of the density distribution at an age
$t_{\rm snr}$ = 3 000 years.}
\end{figure*}

\section{Observations}

After the passage of the reverse shock, the PWN consists of a relic PWN and the
head of the PWN, containing the active pulsar. The figures 7 and 8 show the density 
profiles of the PWN/SNR before and after the formation of the bow shock. For both 
cases the pulsar is located inside the head of the PWN. The synchrotron maps from PWNe 
will therefore show a similar morphology, which makes it hard to determine the presence 
of a bow shock based on the observed PWN/SNR morphology. However, we have demonstrated 
that the following criteria can be used to determine the presence of a bow shock around
the head of the PWN:
\begin{itemize}
\item The ratio between the position of the pulsar $R_{\rm psr}$ with respect to the 
position of the forward shock of the SNR $R_{\rm snr}$ has to satisfy 
$R_{\rm psr}/R_{\rm snr} \ge 0.677$

\item The age of the remnant has to be larger then half the crossing time, i.e. 
$t > 0.5 t_{\rm cr}$.   
\end{itemize}

We will apply the above two criteria to three SNRs: N157B, G327.1-1.1 and W44. We will argue 
that the head of the PWNe inside the first two SNRs are not bounded by a bow shock, and therefore
these PWNe are in either the {\it reverse shock interaction stage} or the 
{\it subsonic expansion stage}. The PWN inside SNR W44 is shown to be the best candidate of a
PWN in the {\it bow shock stage}.


\subsection{The pulsar wind nebula inside N157B}

N157B is a young SNR dominated by plerionic emission from the PWN inside this
remnant. The age of the remnant is approximated to be $5\;000$ yrs (Wang \& 
Gotthelf 1998) and contains a 16 ms pulsar (Marshall et al. 1998). The velocity 
of the pulsar is high ($V_{\rm psr}\simeq 1\;000$ km/sec), if one assumes that the pulsar 
was born at the central region of the bright radio emission (Lazendic et al. 2000). 

Wang \& Gotthelf (1998) argue for a bow shock interpretation of the PWN in N157B:
the supersonic motion of the pulsar has deformed the PWN around the pulsar into a
bow shock. In this scenario the SNR N157B is a young variant of the SNR CTB80, which 
is thought to have a PWN bow shock located close to the shell of the remnant (see e.g. 
Strom 1987; Kulkarni et al. 1988; Migliazzo et al. 2003). It is remarkable
though, that the spindown luminosity of the pulsar inside N157B ($L_0\simeq 4.8\times
10^{38}$ ergs/sec) is so much higher compared with the spindown luminosity from the
pulsar inside CTB80 ($L_0\simeq 4.0\times 10^{36}$ ergs/sec). If there is a bow shock
around the PWN inside N157B, this implies that an upper limit for the age of the SNR,
when the pulsar crosses the shell, is approximately 10 000 years (using the current
age of 5 000 years and the criterion that bow shock creation occurs at half the crossing 
time). This is again a remarkable contrast with the current age of CTB80 of 100 000 years,
which is close to the crossing time of this SNR. Furthermore from figure 2 of Wang et al. 
(2001) it seems like the position of the pulsar is more or less centered in the SNR. This 
is in contrast with what one would expect from the analysis performed in section 2.2 of 
this paper, which predicts $R_{\rm psr}/R_{\rm snr} \ge 0.677$.

\begin{figure*}
\centering
\includegraphics[width=12cm,angle=0]{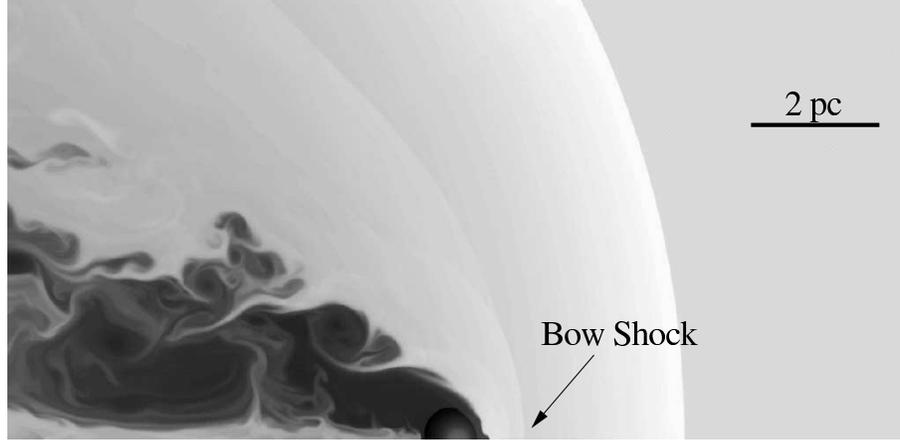}
\caption{Logarithmic gray-scale plot of the density distribution at an age
$t_{\rm snr}$ = 11 400 years.}
\end{figure*}

We propose a scenario for SNR N157B in which the contradictions mentioned above disappear.
Following the results from sections 2 and 3, we suggest that the central bright parts of 
the radio and X-ray emission inside N157B correspond with the relic PWN, being blown in 
the initial stage of the PWN, when it was expanding supersonically. The PWN inside N157B
has just collided with the reverse shock, this collision and the high velocity of the 
pulsar has off-centered the pulsar with respect to the PWN structure as was discussed in the
sections 2 and 3.3 of this paper. Therefore the head of the PWN is not bounded by a bow shock:
the PWN is in the aftermath of the reverse shock interaction stage or in the beginning of the
subsonic expansion stage.


\subsection{The pulsar wind nebula inside G327.1-1.1}

Another example of a young SNR, containing a PWN with the position of the pulsar
off-centered with respect to its PWN is G327.1-1.1. There has not been a pulsar
detected inside this remnant so far. Sun et al. (1999) presented a radio contour 
map using MOST data overlaid with X-ray data from ROSAT. The X-ray emission is 
centered around a finger of radio emission sticking out of a central radio bright 
region, indicating the presence of a pulsar wind. Following Sun et al. (1999) the 
SNR can be modelled in X-rays by the
following set of parameters: $E_{51}=0.23$, $n_0=0.10$, $V_{\rm psr}$=600km/sec and
 an age of $t=1.1\times 10^4$. Using equation (4) to calculate $t_{\rm cr}$ we get 
a value of $4.3\times 10^4$ years. The age of the system therefore is less than half
the crossing time, implying the absence of a bow shock. Furthermore the displacement
of the PWN finger (i.e. the head of the PWN), containing the pulsar, does not satisfy 
the other criterion for the presence of a bow shock, i.e. 
$R_{\rm psr}/R_{\rm snr}\ge 0.677$. This criterion is less restrictive, since one might
introduce an angle between the pulsar velocity and the observer such that the ratio 
$R_{\rm psr}/R_{\rm snr}\ge 0.677$. Notice however that this will not influence the age 
of the remnant!

Therefore we propose a scenario for G327.1-1.1 in which there is no bow shock around the 
head of the PWN: the pulsar has been positioned at he head of the PWN due to the high
velocity pulsar and the passage of the reverse shock. The central bright part of the remnant 
corresponds with the relic PWN. The finger of the PWN corresponds with the head of the
PWN, which contains the active pulsar. Because of the larger value of the ration 
$R_{\rm psr}/R_{\rm snr}$ and the larger age with respect to the PWN inside SNR N157B we
favour the scenario where the PWN is in its subsonic expansion stage.

\subsection{The pulsar wind nebula inside W44}

The SNR W44 is an older remnant then the previous two SNRs discussed. 
Furthermore only a small fraction of the radio emission from this remnant is 
characterised by plerionic emission (Frail et al. 1996). Taking the 
characteristic age of the pulsar, $20\;000$ years (Wolszcan et al. 1991), as 
the age of this remnant yields an upper limit of $t_{\rm cr}\simeq 40\;000$ years 
for the crossing time of this remnant. This age is much closer to
the age of the SNR CTB80. The displacement of the pulsar position only marginally
violates the condition for bow shock formation, i.e. 
$R_{\rm psr}/R_{\rm snr}\ge 0.677$. 
We therefore conclude that the PWN observed in the SNR W44 corresponds with the
head of the PWN, which has been deformed into a bow shock.

\section{Conclusions}

We have considered the case of a PWN interacting with a SNR, for which the 
associated pulsar is moving at a high velocity through the interior of its
SNR. The model we discussed made use of a hydrodynamics code. One could 
distinguish four different stages in the simulation: the supersonic expansion
stage, the reverse shock interaction stage, the subsonic expansion stage and the
bow shock stage. Below we summarise the most important results from our model:
\begin{itemize}
\item
The reverse shock interaction stage starts when the reverse shock collides with
the front of the PWN shock. The timescale on which the reverse shock collides 
with the {\it complete} PWN shock, bounding the pulsar wind bubble, scales with 
the pulsar velocity. This timescale is a significant fraction of the lifetime 
of the PWN, when the reverse shock interaction stage starts, for PWNe containing
a high velocity pulsar.
\item
The high velocity of the pulsar results in an off-centered position of the pulsar 
with respect to its pulsar wind bubble, after the passage of the reverse shock.
\item
The morphology of a PWN, after the passage of the reverse shock, consists of a twofold
structure which ultimately expands subsonically inside the relaxed Sedov-Taylor SNR:
\par
1) a roughly spherically symmetric relic PWN
\par
2) a head, containing the pulsar, directed towards the SNR shell
\item
The formation of a bow shock around the head of the PWN occurs at half the crossing 
time, when the position of the pulsar $R_{\rm psr}$ with respect to the SNR shock 
$R_{\rm snr}$ equals $R_{\rm psr}\simeq 0.677 R_{\rm snr}$. We derived a lower
limit for the pulsar velocity for the bow shock formation to occur while the
SNR is in its Sedov-taylor stage.  
\end{itemize} 
From our model it follows that a SNR containing a pulsar at the head of its PWN
does not imply the presence of a bow shock, bounding the head feature of the PWN. 
We discussed three SNRs and interpreted their morphology, using the results from 
our model. For the SNRs N157B and G327.1-1.1 we 
argued that they contain PWNe which do not have a bow shock. The PWN inside the
SNR W44 seems to be a good candidate for having a bow shock around the head of
its PWN.    
\begin{acknowledgements}
This work was part-funded by the CosmoGrid project, funded under the 
Programme for Research in Third Level Institutions (PRTLI) administered by 
the Irish Higher Education Authority under the National Development Plan 
and with partial support from the European Regional Development Fund.
\end{acknowledgements}

\end{document}